\documentclass{aastex}
\usepackage{spr-astr-addons}

\shorttitle{Far UV and X-ray observation: a hot view of shell galaxies}
\shortauthors{Marino et al.}

\begin{document}

\title{Far UV and X-ray observations: a hot view of shell galaxies}

\author{A. Marino\altaffilmark{1}, R. Rampazzo\altaffilmark{1}, G. Trinchieri\altaffilmark{2}, R. Gr\"utzbauch\altaffilmark{3}, M.S. Clemens\altaffilmark{1}}
\affil{{1} INAF - Osservatorio Astronomico di Padova, Vicolo dell'Osservatorio 5, 
I-35122 Padova, Italy}                                                         
\affil{{2} INAF - Osservatorio Astronomico di Brera, Via Brera 28, I-20121 Milano, Italy}
\affil{{3} Institut f\"ur Astronomie der Universit\"at Wien, T\"urkenschatzstrasse 17, 1180 Wien, Austria}

\begin{abstract}
Shell galaxies are considered the debris of recent accretion/merging episodes.
Their high frequency in low density environments suggest that such episodes could drive the secular evolution for  at least some fraction of the early-type galaxy population. 
We present here the preliminary results of ultraviolet and X-ray data for a sample of  three shell galaxies, namely NGC~474, NGC~7070A and ESO~2400100.  The Far UV morphology and photometry are derived 
using the  observations obtained with the {\it Galaxy Evolution Explorer} and the XMM-{\it Newton} Optical Monitor. We aim at investigating the {\it rejuvenation} processes in the stellar population using the UV information as well as at gaining information about the
possible evolution with time of the X-ray emission due interaction/merging 
processes. 

\end{abstract}


\section{Introduction}
In a hierarchical evolutionary scenario, galaxies experience accretion/merging
events during their lifetime. While early-type  galaxies in nearby clusters appear
(homogeneously) old, the field early-type galaxy population seems to contain 
genuinely, recently {\it rejuvenated} objects \citep[see e.g.][]{Clemens06}.
Early-type galaxies showing fine structure, like shells, occupy a special position
 since they are believed to fill the gap between ongoing mergers and normal elliptical galaxies.

Shells are faint, sharp-edged stellar features  characterizing $\approx$ 16.5\% 
of the field early-type galaxy population and avoiding  cluster environments 
\citep[e.g.][]{ Malin83, Reduzzi96, Colbert01}. 
Two major scenarios for their origin have been proposed: {\em a)} a weak interaction 
between galaxies \citep{Thomson90, Thomson91}, {\em b)} merging/accretion events 
\citep{Dupraz86, Hernquist87a, Hernquist87b}. In the former scenario, weak interaction
 can form long lasting azimuthally distributed shells through the interference of density waves 
producing a thick disc population of dynamically cold stars.  However, this 
requires a cold thick disc, not found  in ellipticals. In the  merging/accretion events 
between galaxies of different masses (mass ratios typically 1/10 - 1/100, {\em b)} scenario above) 
 shells are density waves formed from infalling stars from the companion during a minor merger.
 Major merger can also  produce shells \citep{Barnes92, Hernquist92, Hernquist95}.  Whatever the mechanism is, interaction/merging events seem to have played a significant role in the formation/evolution of the early-type class
in the field and  activated a new star formation event 
\citep{Longhetti99,Longhetti00}.
 
Whether there is a link between shell galaxies, the early phases of merging processes, and the class of 'normal' early-type galaxies 
 is still an open question. 
To test the depth of the link, 
 a multiwavelength approach is of paramount importance. The UV  data probe the ongoing/recent  
  star formation processes and study their distribution across the galaxy while 
the X-ray emission is 
 connected to the past star formation and metal enrichment history  of the 
 bulk of the galaxy and may disclose hidden AGN activity.

We present  {\it Galaxy Evolution Explorer} ({\it GALEX}), XMM-{\it Newton} Optical Monitor (OM) and XMM-EPIC observations of  NGC 474, NGC 7070A, and ESO 2400100, three shell galaxies in the \citet{Malin83} catalogue.

\begin{table*}
\tabletypesize{\scriptsize}
\caption{Journal of the XMM-{\it Newton} EPIC and  Optical Monitor observations}
\begin{tabular}{llllllllll}
 \hline
 \multicolumn{1}{l}{Galaxy}& 
 \multicolumn{1}{c}{EPIC}  &
 \multicolumn{1}{c}{EPIC}  &
\multicolumn{1}{c}{OM}  &
\multicolumn{1}{c}{OM}  & 
\multicolumn{1}{c}{OM}  &
\multicolumn{1}{c}{OM}  &
\multicolumn{1}{l}{observing} &
\multicolumn{1}{l}{observation}\\
 \multicolumn{1}{l}{}& 
 \multicolumn{1}{c}{PN}  &
 \multicolumn{1}{c}{MOS}  &
\multicolumn{1}{c}{UVM2}  &
\multicolumn{1}{c}{UVW1}  & 
\multicolumn{1}{c}{U}  &
\multicolumn{1}{c}{B}  &
\multicolumn{1}{l}{date} &
 \multicolumn{1}{l}{ident.}\\
\multicolumn{1}{c}{}&
\multicolumn{1}{c}{ [sec]} &
\multicolumn{1}{c}{ [sec]} &
\multicolumn{1}{c}{ [sec]} &
\multicolumn{1}{c}{ [sec]} &
\multicolumn{1}{c}{ [sec]} &
\multicolumn{1}{c}{ [sec]} &
\multicolumn{1}{c} {}&
\multicolumn{1}{c}{}\\
\hline
NGC~474        &   4300 & 11400  &   5000 &  5000  &  5000 &  & 2004-01-24 & 
0200780101 \\
NGC~7070A    &  26250 & 30840  &  4400  &          &         & 4000  &2004-10-28  & 0200780301  \\
ESO 2400100  & 19970  & 26330  & 4400    &  4400  & 4400 &  &2004-05-11  & 0200780201 \\
\hline
\end{tabular}

\medskip
{XMM-{\it Newton} observations of NGC 474 are fully discussed in \citet{Rampazzo06} }
\label{table1}
\end{table*}

\section{The sample}

Beside the presence of shells, clearly visible in all three, although more spectacular in NGC 474, 
these galaxies share other properties:  
they are in  low density environments  and
are  interacting systems. However, the details of the  interaction are 
quite different: companions are well separated in  NGC 7070A 
\citep{Ramella96} and strongly interacting in  NGC 474 as shown by H{\sl I} 
observations \citep[see e.g.][]{Rampazzo06}. Two nuclei are embedded and interacting within the  ESO 2400100 envelope \citep{Longhetti98b}. 
Furthermore, dust is detected in  NGC~474 and NGC~7070A, although only for the latter, there is
 kinematic evidence of an external acquisition.

\begin{figure}
\resizebox{8cm}{!}{
\includegraphics[width=5cm]{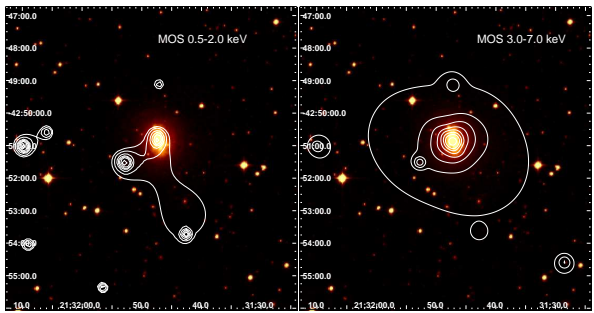}}
\caption{Isointensity contours from adaptively smoothed XMM-Newton combined MOS images in two bands on DSS2 plate for NGC 7070A (left panels) and ESO2400100 (right panels).}
\label{fig1}
\end{figure}

\section{Observations and reduction}

Table~\ref{table1} gives the Journal of the  XMM-Newton
observations.
X-ray data were treated with the standard routines provided by  
SAS version 7.0 
as suggested by XMM-Newton 
Science Analysis system:Users' Guide available 
on line\footnote{\tt http://xmm.vilspa.esa.es/external/xmm \_user\_support/
documentation/sas\_usg/USG}.
The  X-ray observations of  NGC 474 have been already  discussed in \citet{Rampazzo06}. 

During X-ray observations,  shell galaxies were simultaneously
imaged in the ultraviolet and optical bands with the Optical Monitor 
\citep{Mason01}).  Observations have been performed using UVW1 and UVM2
bands which cover the ranges 245-320 nm and 205-245 nm, respectively. 
Galaxies have been observed also in the U (300-390 nm)
and B (390-490 nm) bands. The Point Spread Function -- FWHM -- is 
$\approx$2.0\arcsec\ in UVW1, and 1.8\arcsec\ in UVM2 sampled
with 0\farcs 476$\times$0\farcs 476 pixels. 

In order to complete the UV information about our galaxies we
searched the {\it GALEX} archive and retrieved the data for NGC 474.
The {\it GALEX} mission and instruments are fully described  in
\citet{Martin05} and \citet{Morrissey05}.  The spatial resolution 
of the images is $\approx$4\farcs 5 and 6\farcs 0 FWHM in FUV 
(135-175 nm) and NUV (175-275 nm) respectively, 
sampled with 1\farcs 5$\times$1\farcs 5 pixels.
We found {\it GALEX} data only for the NGC~474, that  was
observed the October 5th, 2003 with an exposure of 1477 and 1647 
seconds in NUV and FUV bands respectively.

\begin{figure*}
    \resizebox{17cm}{!}{
\includegraphics[width=5cm]{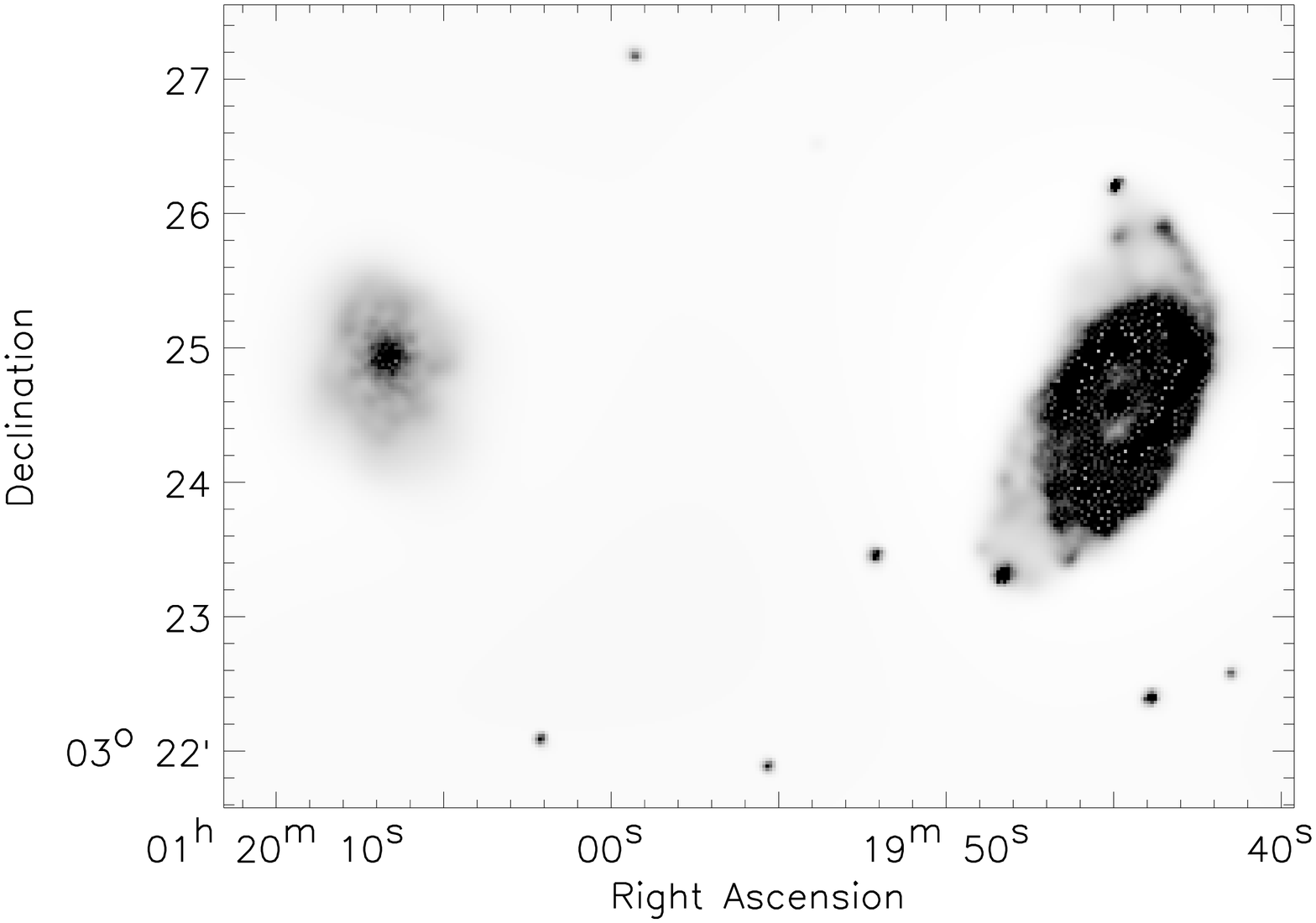} 
\includegraphics[width=5cm]{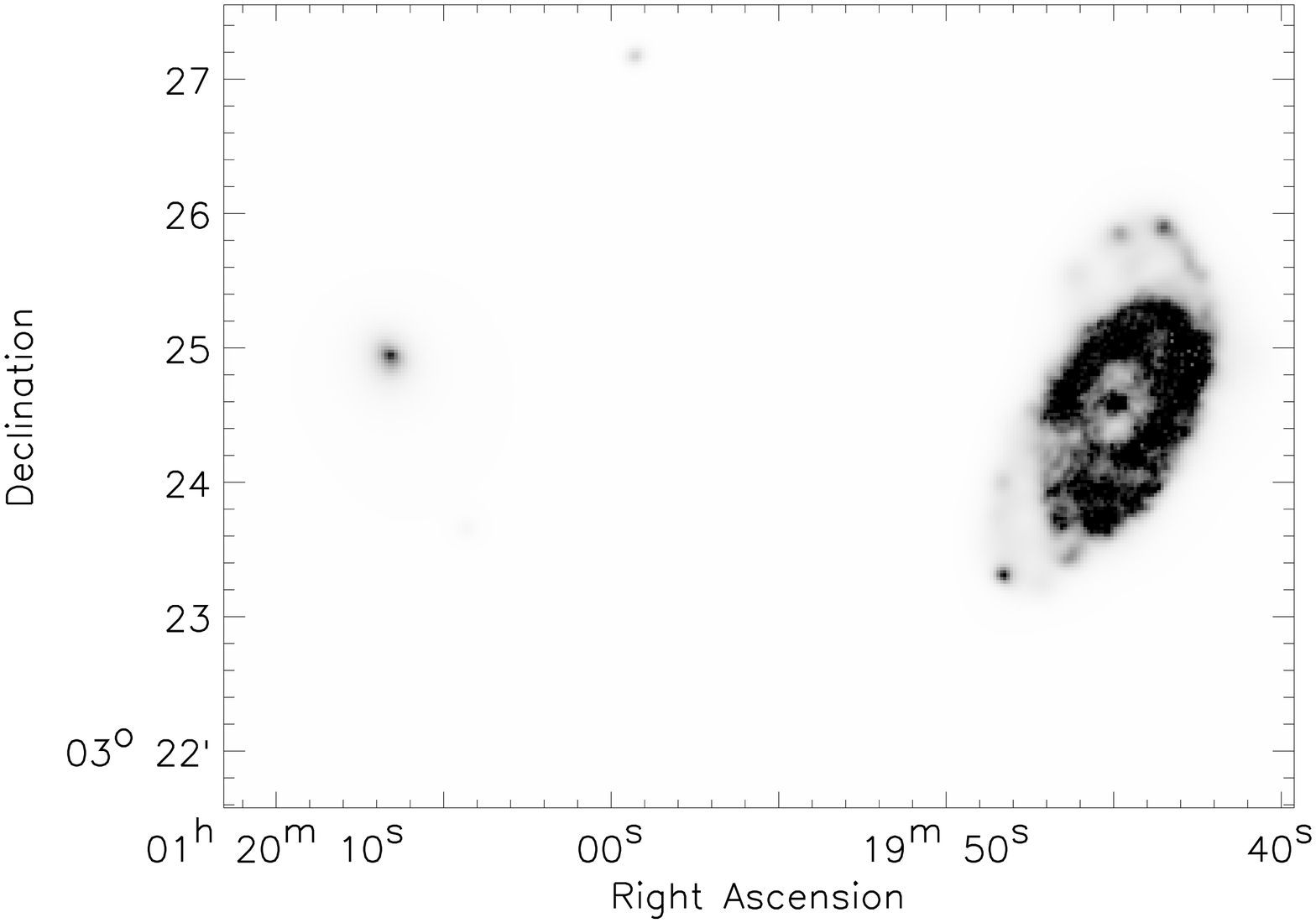}
\includegraphics[width=5cm]{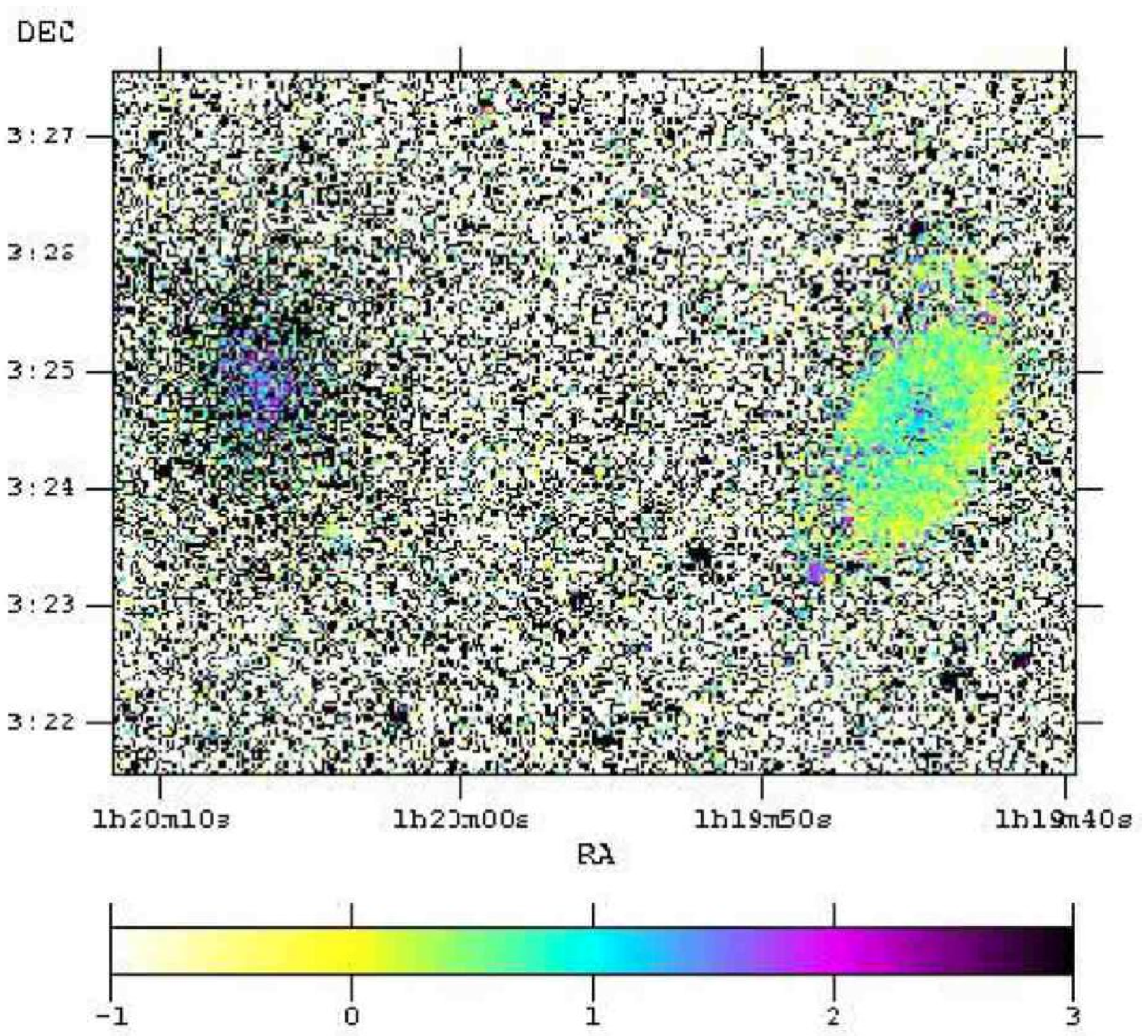}}
\caption{The NGC 470/474 system, also known as Arp 227. Full resolution  {\it GALEX} FUV (left panel) and NUV (mid panel) background subtracted images. 
Pixel by pixel  GALEX (FUV-NUV) 2D color map of the NGC 470/474
system (right panel). The (FUV-NUV) color of both the early-type type galaxy, NGC 474, and of the spiral, NGC 470, are typical of their class.
Most likely, the
NUV and the FUV fluxes have different origins in `normal' early-type galaxies.
The NUV flux is partially due to the MS turn-off stars of the evolved population
and partially to more evolved and, exotic stars. In contrast, a strong FUV emission
is likely due to the presence of one (or more) hot, plausibly high metallicity,
stellar component giving origin to the well known phenomenon of the 
UV-upturn (e.g. hot-HB and/or post-AGB and AGB-manqu\' e stars).  In shell galaxies, the presence 
of young stars could also contribute in different proportion, to both NUV and FUV fluxes.}
\label{fig2}
\end{figure*}

\begin{figure*}
    \resizebox{17cm}{!}{
\includegraphics[angle=-90,width=6.5cm]{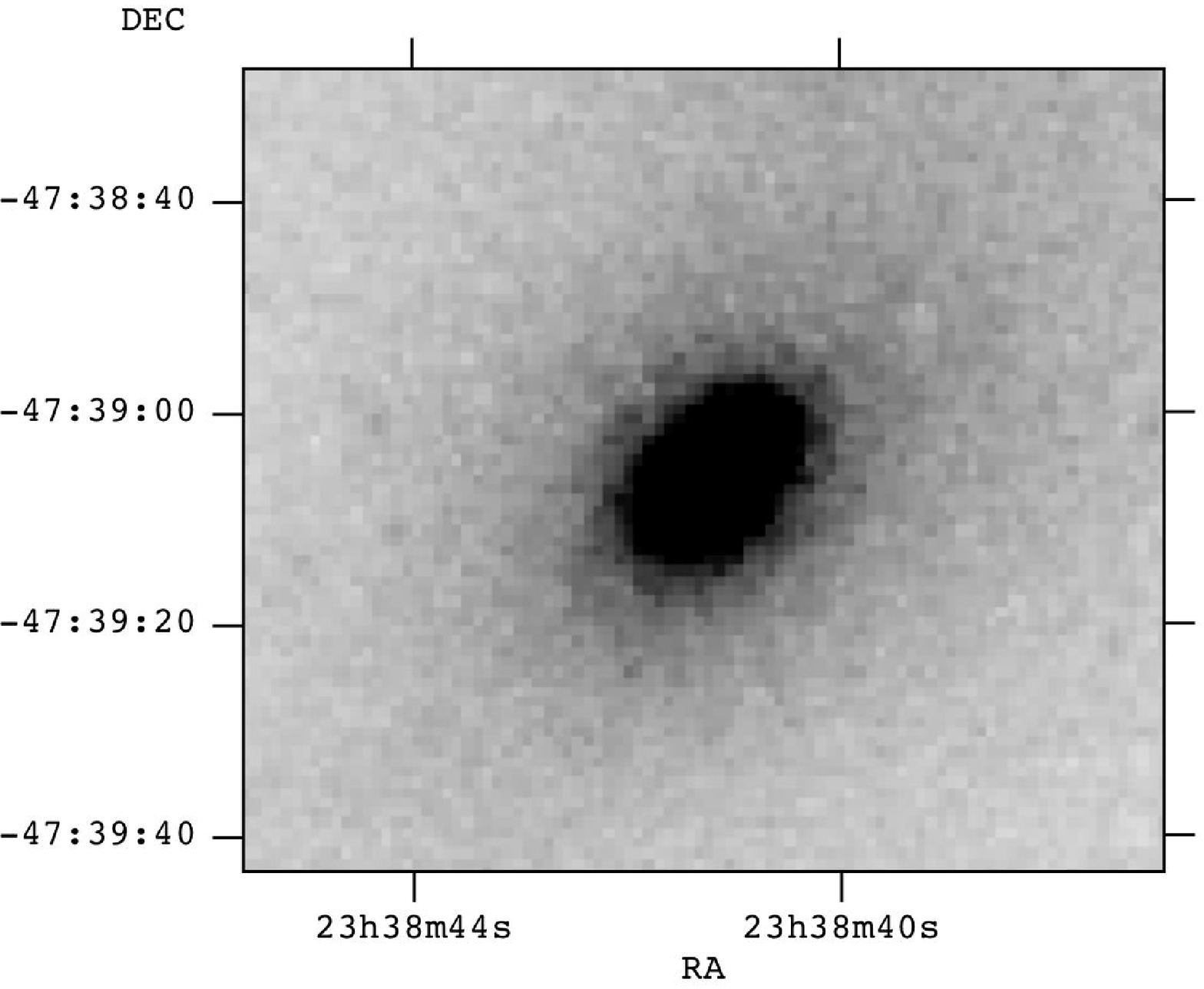}
   \includegraphics[angle=-90,width=6.5cm]{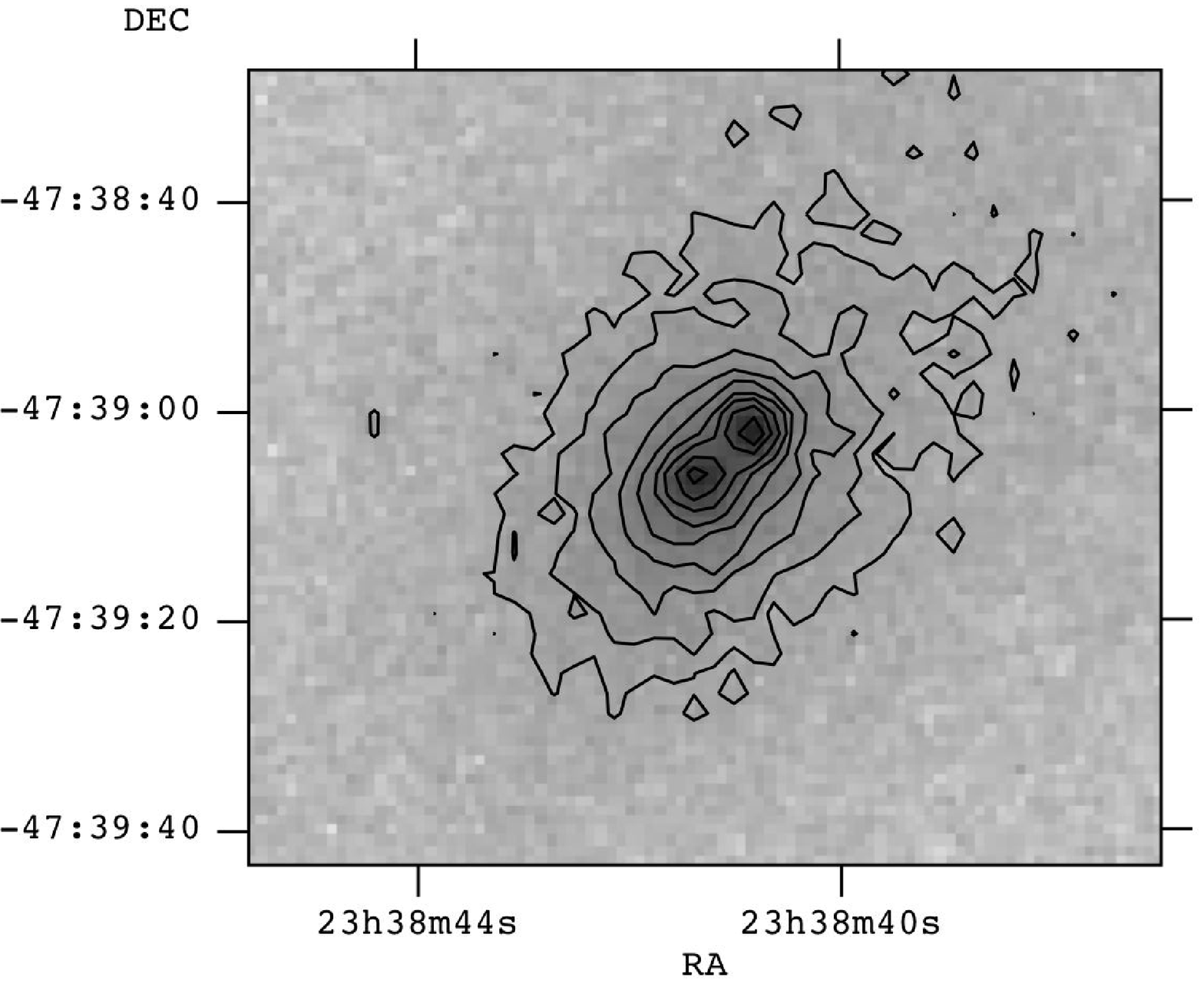}
   \includegraphics[angle=-90,width=6.5cm]{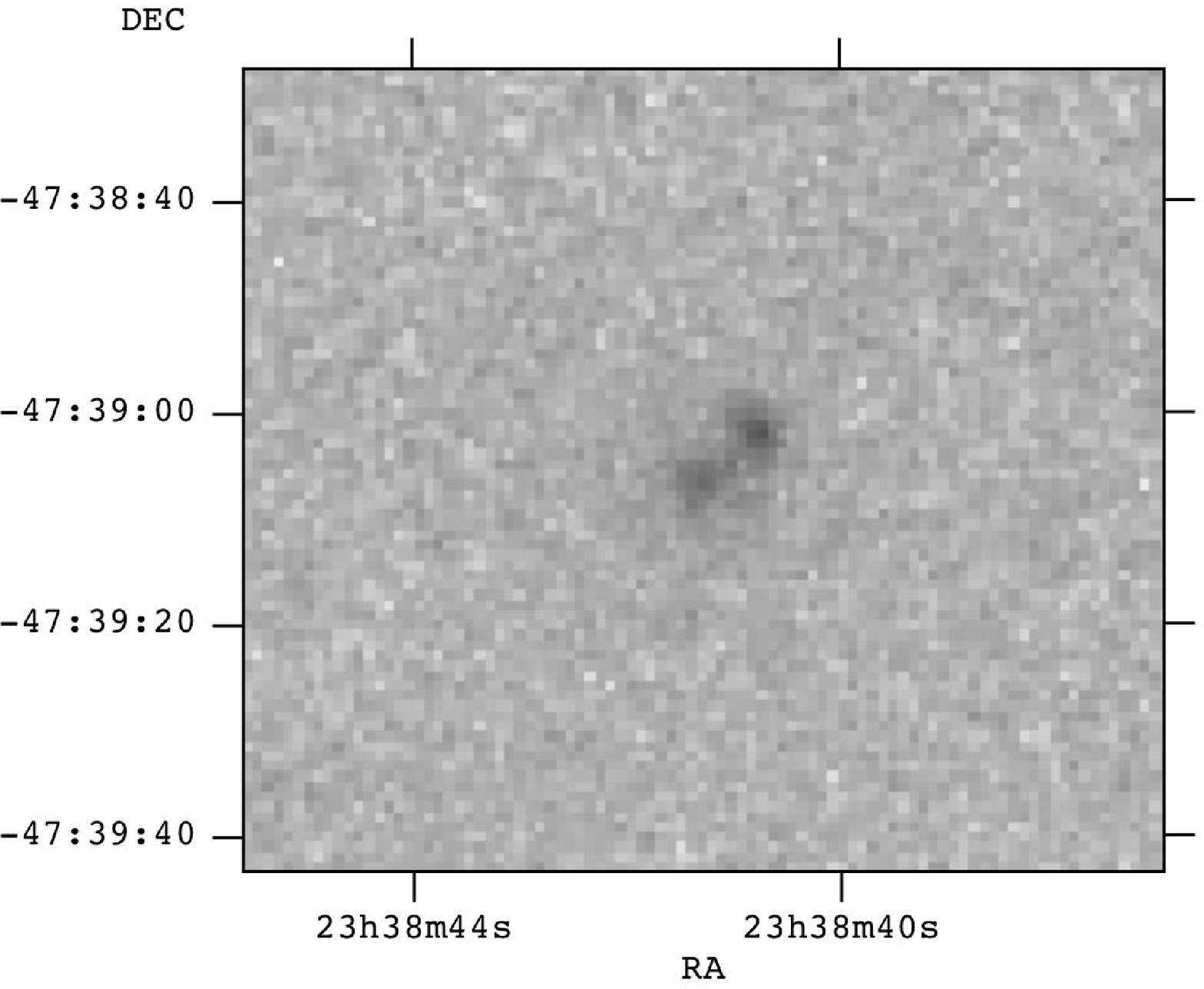}}
    \caption{U (left panel), UVW1 (mid panel) and UVM2 (right panel) 
    images of ESO~2400100. The two nuclei (we named ESO~2400100$a$ the 
    southern, ES0~2400100$b$ the northern)  embedded in the galaxy body 
    are clearly visible in the UVW1 and UVM2 images.}
\label{fig3}
\end{figure*}

\section{Preliminary results}

We aim at obtaining crucial 
 information about  the time at which the accretion/merging phenomenon has occured that can be derived from the far UV and optical colours.
Such information is  important  for a proper
discussion of the X-ray data. According to the evidence provided by the sample of \citet{Brassington07}, the X-ray luminosity could  evolve during the period
that characterizes phases of the galaxy-galaxy encounter, the merging and the 
set-up of the merging towards a relaxed galaxy. 

Figure~\ref{fig1} shows the isointensity contours from the adaptively smoothed 
XMM-{\it Newton} of NGC 7070A and ESO 2400100 superposed on the DSS plates. 
It is evident that the emission from these two galaxies has a significantly 
different morphology. In NGC 7070A  
the emission is rather compact both in the hard and soft bands. In 
ESO 2400100 the soft band emission is significantly stronger than the hard
one and  extends  further than in NGC 7070A. Spectral results are in agreement 
with the spatial picture: the nuclear source of NGC 7070A is the dominant 
component while in ESO 2400100 the emission is indicative of a low 
temperature plasma. \citet{Rampazzo06} showed that NGC 474 is at the bottom
of the X-ray emission distribution of E galaxies: its X-ray emission is consistent
with the low end of the expected emission from discrete sources.

{\it GALEX} data of NGC~474 (Figure~\ref{fig2})  show that the NUV emission extends to the galaxy bulge,  while the FUV emission shows up only 
in the central regions of the galaxy. In the NGC 474 image obtained in 
the UVW1 and UVM2 filters outer shells are visible as in the optical image 
\citep{Rampazzo06}.    
Both ESO~2400100 (Figure~\ref{fig3})
and NGC~7070A have extension similar to that of the optical image in the OM-UVW1  
and in  the {\it GALEX} NUV  filter.
This implies that the UV emission comes from the same kind of stellar
population. The FUV emission, 
more concentrated toward the nucleus is most likely dominated by the emission of different types of
hot stars \citep[see also][]{Rampazzo07}.   

In Figure~\ref{fig3} (mid and left panels) the UVW1 and UVM2 images 
show the two nuclei embedded in ESO 2400100. The northern nucleus is
significantly bluer than the southern one. \citet{Longhetti00} noticed that the
two nuclei have a significantly different H$\beta$ line-strength indices
(2.79 in the northern nucleus vs. 1.54 in the southern one) suggesting a different stellar population composition.

The accretion of faint galaxies seems one of the drivers of the secular
evolution of galaxies in loose, poor groups. Since shells are widely believed to be generated by an accretion event, shell galaxies potentially trace the typical secular evolution  of early-type galaxies in such environments.  Furthermore, there  is a growing evidence of multiple accretion events in the same galaxy. The complex shell system of NGC 474 is believed to be generated by  two distinct accretion events \citep{Sikkema07}. In this framework the double nucleus in ESO~2400100 is an interesting and puzzling case at the same time. 
Is the second nucleus the cause of the observed system of shells or another evidence of a new, on-going, accretion event?

\acknowledgments
This research has been partially founded by ASI-INAF contract I/023/05/0.
Galex is a NASA Small Explorer, operated for NASA by California Institute of 
technology under NASA contract NAS-98034.

\end{document}